\documentstyle[preprint,aps]{revtex}
\begin{document}
\input psfig


\preprint{UTPT-95-17}

\title{Galaxy Dynamics Predictions in the Nonsymmetric Gravitational Theory}

\author{J. W. Moffat and I. Yu. Sokolov}

\address{Department of Physics, University of Toronto,
Toronto, Ontario, Canada M5S 1A7}

\date{\today}

\maketitle

\begin{abstract}
In the weak field approximation, the nonsymmetric gravitational theory
(NGT) has, in addition to the Newtonian gravitational potential,
a Yukawa-like potential produced by the exchange of
a spin $1^+$ boson between fermions. If the range $r_0=\mu^{-1}$ is
$25$ kpc, then this additional potential due to the interaction with matter
in the halos
of galaxies can explain the flat rotation curves of galaxies and the
Tully-Fisher law ($L\sim v^4$) without the dark matter hypothesis. Possible
fits to clusters of galaxies and gravitational lensing observations are
discussed. The results are based on a linear approximation to a new
perturbatively consistent version of the NGT field equations, which does
not violate the weak equivalence principle.
\end{abstract}

\pacs{ }

\narrowtext

\section{Introduction}

After two decades there has not been any observation of
exotic dark matter candidates. Recent observational results using the HST
have excluded faint stars as a significant source of dark matter in the solar
neighborhood\cite{Hubble}. However, the galaxy dynamics observations
continue to
pose a serious challenge to gravitational theories. The data are in sharp
contradiction with Newtonian dynamics, for virtually all spiral galaxies
have rotational velocity curves which tend towards a constant
value\cite{Kent,Aaronson,Sanders1}.

As in the case of anomaly problems in solar dynamics of the past century,
concerning Uranus and Mercury, there are two ways to circumvent the
problem. The most popular is to postulate the existence of dark matter
\cite{Ashman}.
It is assumed that dark matter exists in massive almost spherical halos
surrounding galaxies. About 90\% of the mass is in the form of dark matter
and this can explain the flat rotational velocity curves of galaxies.
However, the scheme is not economical, because it requires three or more
parameters to describe different kinds of galactic systems and no satisfactory
model of galactic halos is known.

The other possible explanation for the galactic observations is to say that
Newtonian gravity is not valid at galactic scales. This has been the
subject of discussion in recent years\cite{Bekenstein}-\cite{Nester}.
We know that Einstein's
gravitational theory (EGT) correctly describes solar system observations
and the observations of the binary pulsar PSR 1913+16\cite{Will}.
Therefore, any explanation of galactic dynamics based on gravity must be
contained in a modified gravitational theory that is consistent with EGT.
The following constraints on a classical gravitational theory are:

{\smallskip\obeylines
(1) The theory must be generally covariant, i.e., the field equations should
be independent of general coordinate transformations and should reduce to
special relativity dynamics in flat Minkowskian spacetime.

(2) The theory should be derivable from a least action principle in order to
guarantee the consistency of the theory.

(3) The linear approximation should be consistent, i.e., there should not be
any ghost poles, tachyons or higher-order poles and the asymptotic flat space
boundary conditions should be satisfied.

(4) The equations of motion of test particles should be consistent with local
equivalence principle tests.

(5) All solar system tests of gravity and the observed rate of decay of the
binary pulsar should be predicted by the theory.\smallskip}

We shall now consider the predictions for galaxy dynamics in a new version
of the nonsymmetric
gravitational theory which can satisfy all the above
criteria\cite{Moffat1,Moffat2,LegareMoff1}.
The theory has a linear approximation free of ghost poles, tachyons and
higher-order poles with field equations for a massive spin $1^+$ boson with
a range parameter, $r_0=\mu^{-1}$, corresponding to Proca-type equations for an
antisymmetric potential. The expansion of the field equations about an
arbitrary EGT background metric is also consistent and satisfies the
physical boundary conditions at asymptotically flat infinity.

\section{Test Particle Acceleration in NGT}

A derivation of the equations of motion of test particles yields
the following additional NGT test particle radial acceleration in the
weak field approximation\cite{LegareMoff2}:
\begin{equation}
\label{facceleration}
a_{\hbox{ngt}}(r)=\frac{\lambda\gamma(r)c^2}{\alpha(r)}
\biggl(1-\frac{\lambda\gamma(r) f'(r)}
{\sqrt{r^4+f^2(r)}}\biggr)^{-1}\frac{d}{dr}\biggl(\frac{\gamma(r) f'(r)}
{\sqrt{r^4+f^2(r)}}\biggr),
\end{equation}
where $r$ is the radial distance from a point source, $\lambda$ is a
coupling constant, $c$ is the velocity of light, and
\begin{equation}
\gamma(r)\approx\alpha(r)^{-1}=1-\frac{2GM}{c^2r}.
\end{equation}
Moreover, $f(r)$ is defined by
\begin{equation}
g_{[23]}(r)=f(r)\sin\theta,
\end{equation}
where $g_{[23]}$ is the only non-zero component of the antisymmetric part
$g_{[\mu\nu]}$ of the nonsymmetric tensor $g_{\mu\nu}$, defined by
$g_{\mu\nu}=g_{(\mu\nu)}+g_{[\mu\nu]}$.

For  $r \geq 0.2$ kpc (see Section VI), we have\cite{LegareMoff2}:
\begin{equation}
\label{f1equation}
f(r)\approx C\exp(-r/r_0)(1+r/r_0),
\end{equation}
where $C$ is a constant of dimension $(\hbox{length})^2$. Substituting
(\ref{f1equation}) into (\ref{facceleration}), we get the total radial
acceleration experienced by a test particle in a static spherically symmetric
gravitational field:
\begin{equation}
\label{totalaccel}
a(r)=-\frac{GM}{r^2}+\frac{\lambda
Cc^2}{r_0^2}\frac{\exp(-r/r_0)}{r^2}(1+r/r_0).
\end{equation}

For $r \ll r_0$, we find that\cite{LegareMoff2}:
\begin{equation}
\label{f2equation}
f(r)\approx C\biggl[1-\frac{1}{2}\biggl(\frac{r}{r_0}\biggr)^2
+\frac{2GM}{c^2r}\biggr].
\end{equation}
Substituting this expression into (\ref{facceleration}), we obtain the
additional NGT acceleration:
\begin{equation}
\label{addacceleration}
a_{\hbox{ngt}}(r)=\frac{4C\lambda GM(C^2+2r^4)}{r^3(C^2+r^4)^{3/2}}.
\end{equation}

In the new version of NGT, the weak equivalence principle is satisfied
for the motion of test particles, whereas the strong equivalence
principle is not satisfied. Thus, test particles fall in a local
gravitational field independently of their composition, while the
non-gravitational laws of physics in local Minkowskian frames are
not equivalent to one another. We therefore expect to satisfy all
the E\"otv\"os-type experiments for which bodies with baryon or lepton
number charges fall at the same rates in a gravitational
field\cite{Will,Stubbs,Faller}.

We shall apply these formulas to try to explain the observed long-standing
paradox of the flatness of the rotation curves of galaxies, as well as
the Tully-Fisher law\cite{Tully}, without postulating large amounts of dark
matter in galaxies.

The above formulas will only be used for the acceleration due to points
sources.
A more extensive solution requires, for example, an integration over
a disk density profile. We shall not attempt to do such a fit in the
following, but leave this problem for a future publication.

\section{Fitting the Rotation Curves}

As one can see from the previous section, there are two free parameters,
$\lambda$ and $r_0$, and a constant $C$ that remains to be fixed. We shall
choose the constant $C$ to be
\begin{equation}
C\propto\sqrt{M}.
\end{equation}
We then get the total radial acceleration on a test particle:
\begin{equation}
a(r)=-\frac{G_{\infty}M}{r^2}+\sigma\sqrt{M}\frac{\exp(-r/r_0)}{r^2}
(1+r/r_0),
\end{equation}
where $\sigma$ is a constant, $G_{\infty}$ is defined to be the gravitational
constant at infinity. We are assuming that the gravitational constant can be
different at small and large distance scales. We shall also assume that
the range parameter $r_0$ is compatible with galaxy scales.

Let us assume that Newton's law of gravity is valid with the standard
gravitational constant, $G_0=6.673\times 10^{-11}\, N/kg/m^2$, for
distances that are smaller than the galaxy scale $\sim 10$ kpc. It is
known that the deviations of the rotation curves of spiral galaxies from the
simple Newtonian law (if we do not assume the dark matter hypothesis),
take place for distances starting at $\sim$ 1-4 kpc, depending on the
galaxy mass. For the sake of simplicity, let us put $\sigma=G_0\sqrt{M_0}$,
where $M_0$ is a constant mass parameter. In order to guarantee that we
obtain the usual Newtonian law for sufficiently small $r$, we set
\begin{equation}
\label{renormgrav}
G_{\infty}=G_0\biggl(1+\sqrt{\frac{M_0}{M}}\biggr).
\end{equation}
We then get the expression for the total radial test particle acceleration:
\begin{equation}
\label{finalacceleration}
a(r)=-\frac{G_0M}{r^2}\biggl\{1+\sqrt{\frac{M_0}{M}}[1-\exp(-r/r_0)
(1+r/r_0)]\biggr\}.
\end{equation}

We can now find the dependence of the rotation velocity on the mass $M$
and the distance $r$ by using the formula: $a=v^2/r$. We have the
two free constants $M_0$ and $r_0$ to fit the rotation curves of galaxies.
Since our formulas are valid only for point mass sources, it means that
we can fit only velocities of stars and hydrogen gas that rotate quite
far from the disk of galaxy matter. Fortunately, the problem of the flatness
of the rotation curves is associated with this region of the galaxies.
For distances less than $1-4$ kpc, the standard Newtonian law of gravity
will apply.

Taking the information about rotation curves from Refs.\cite{Kent,Aaronson},
we have found at least 9 spiral galaxies in which the rotation curves
were measured far from the disk of galaxy matter. They are: N2403,
N2903, N3109, N3198, N5033, N5055, U2259, N3031 and N7331. The flat
8regions of the dependence of rotation velocities on the distance
from the galaxy center (plateaus) is $\sim$ 10 to 40 kpc, so
Eq.(\ref{finalacceleration}) must generate a flat function of $r$ for
such distances. A similar problem was solved by Sanders\cite{Sanders3}
for one of the variants of modified Newtonian dynamics (MOND).
In analogy, we find for the case of Eq.(\ref{finalacceleration}) that
it yields flat curves if $r_0\sim 20$ kpc and $5 < \sqrt{M_0/M} < 100$.

In the actual comparison with observations, we shall parameterize the
velocity not by the mass $M$ but by the luminosity $L$, which is proportional
to the mass. We shall use $L/M\sim 3$ and this gives for the rotational
velocity:
\begin{equation}
\label{velocity}
v^2=\frac{3G_0L}{r}\biggl\{1+\sqrt{\frac{L_0}{L}}[1-\exp(-r/r_0)
(1+r/r_0)]\biggr\},
\end{equation}
where $L_0=3M_0$.

Fig.1 shows an example of fitted curves for three galaxies from the above
list for which there were enough data (at least six points to plot).
For the best fit, we found the
parameters to be: $L_0=250\times 10 ^{10}L_{\odot}$ and $r_0=25$ kpc.
In these fits we treated $L$ as a free parameter for all the galaxies.
{}From the fits we obtained a best set of $L$'s for the galaxies, which
we called $L_{\hbox{theor}}$. In
Fig.2, we display the dependence of $L_{\hbox{theor}}$ on the observed
$L_{\hbox{observ}}$, plotted for all the nine galaxies we have considered.
The solid line corresponds to $L_{\hbox{theor}}=L_{\hbox{observ}}$. As
we can see, the agreement is very good. The scattering of the results around

\begin{figure}
\psfig{figure=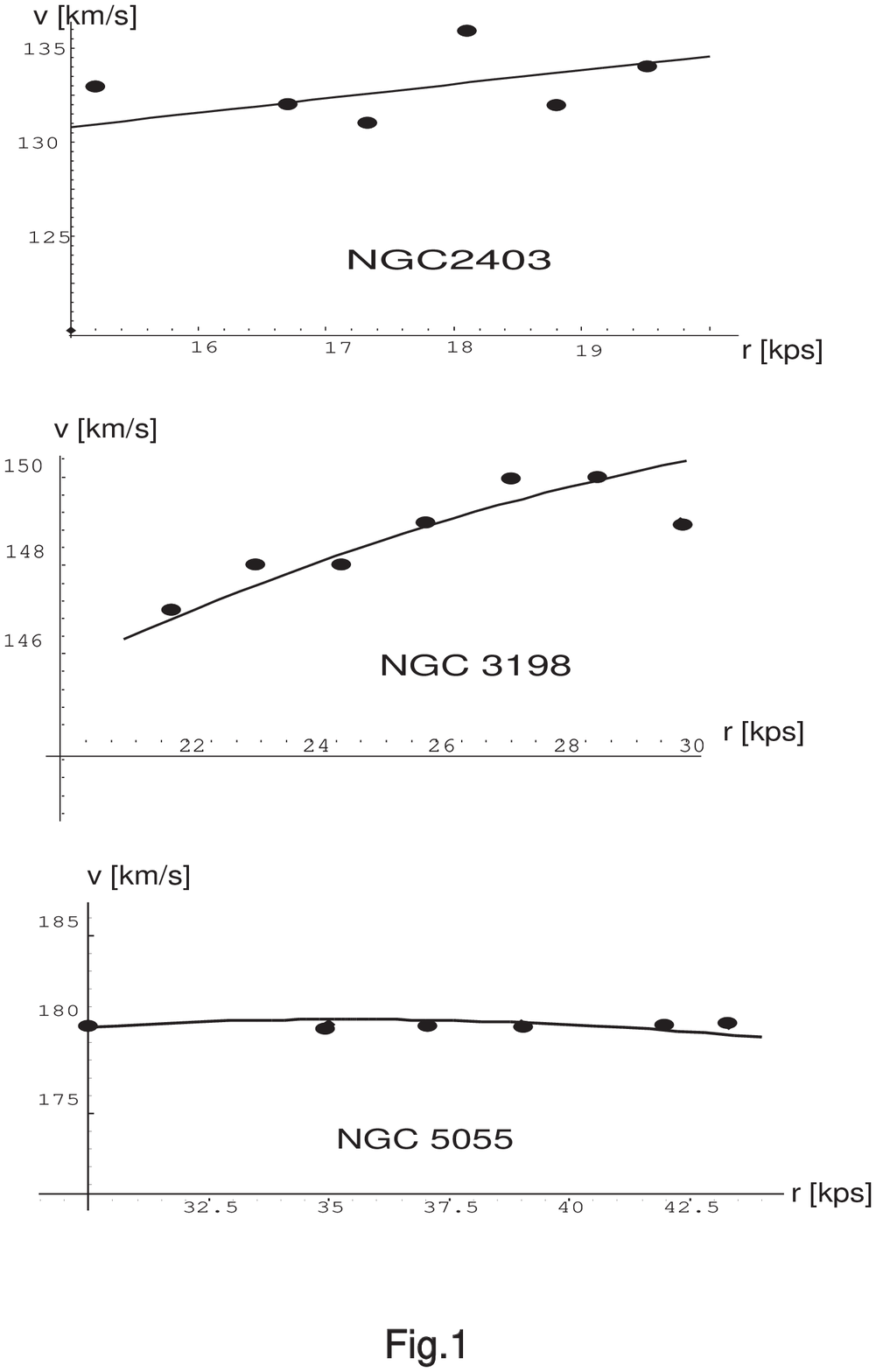,width=6in}
\end{figure}

\noindent the line $L_{\hbox{theor}}=L_{\hbox{observ}}$ is normal for this kind
of
data and can be explained by the measurement errors and by the existence
of some fraction of dark baryonic matter.

\begin{figure}
\centerline{\psfig{figure=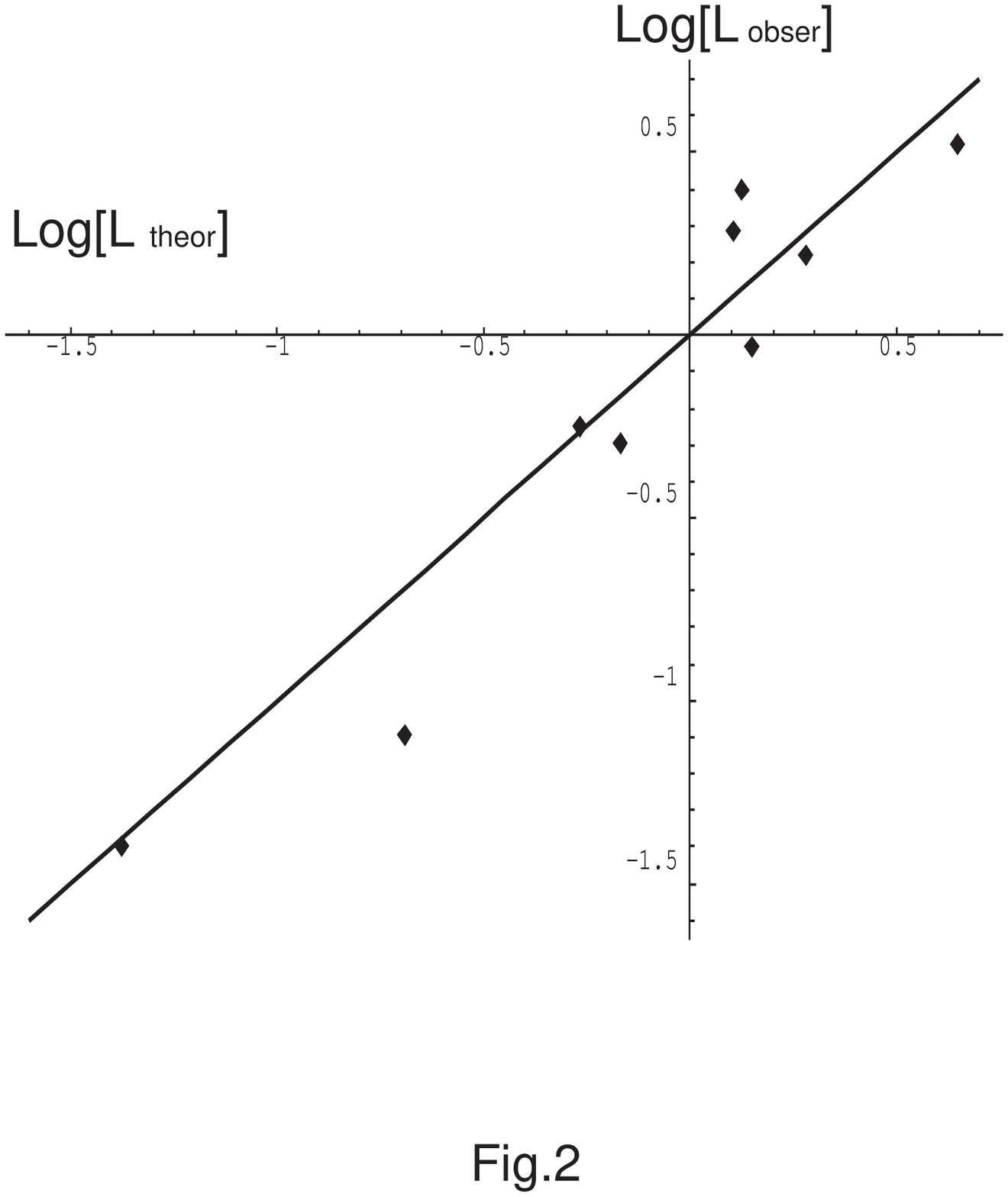,width=4in}}
\end{figure}

\section{The Tully-Fisher Law}

Eq.(\ref{velocity}) gives a good fit to the Tully-Fisher law, for which
$L^\beta\sim v^4$ where $\beta=1$. Fig.3 presents a graph of the value of
the parameter $\beta$ as a function of $r$ and the absolute magnitude $H$,
defined by $H=-2.5\log(L/L_{\odot})+ b$, where $b$ is a constant.
We choose $b=0$ and consider the interval of $H$ corresponding to
$10^8 < L/L_{\odot} < 5\times 10^{10}$, which covers all the galaxies
considered here. The case $\beta=1$ corresponds to the exact
Tully-Fisher law, if we put the maximum difference between radial
velocities in a galaxy equal to $2v$ (the difference between
the forward and backward velocities). In the same figure, we show the plane
$\beta=1.1$. If we take into account that bigger $r$ corresponds to
bigger $L$ (bigger absolute value of $H$), which means that the larger
galaxies are more luminous, we see from the figure that the deviation
from $\beta=1$, i.e. the Tully-Fisher law is about $10-15 \%$.

\begin{figure}
\centerline{\psfig{figure=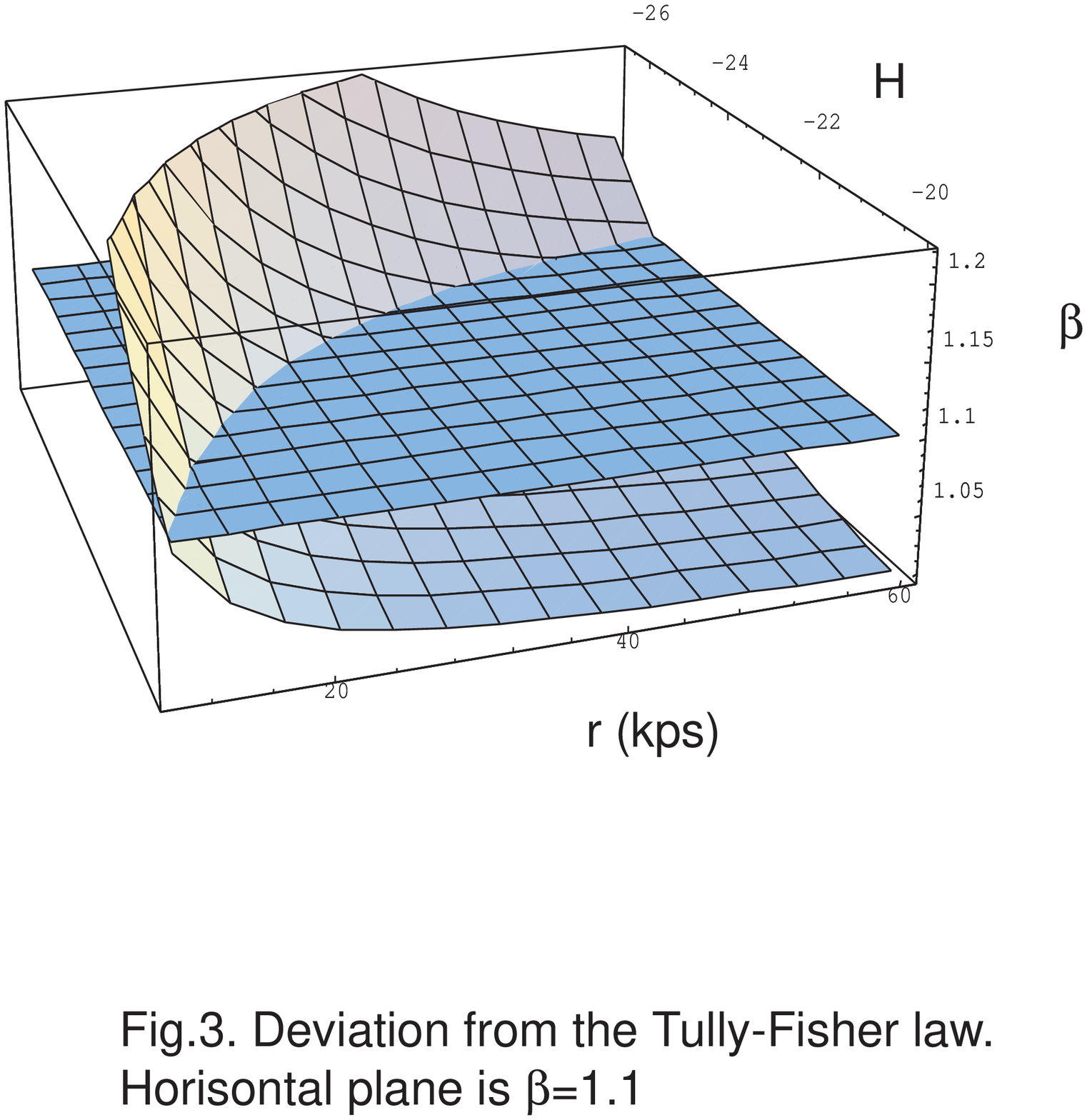,width=4in}}
\end{figure}

We recall that we have used the point mass approximation for the galaxies.
Presumably in an exact calculation, the repulsive NGT force would give a
larger contribution\cite{Sokolov}, which means that
an exact calculation would give a diminished radial velocity for a
given $L$. This would result in a decrease of $\beta$.

Let us now consider the approximation that the maximum difference between
radial velocities in a galaxy is equal to $2v$. From Fig.4, we see that
for large galaxies with a big luminosity (mass) ($L > 5\times 10^{10}
L_{\odot}$) the maximum velocity is bigger than the plateau velocity, whereas
for small galaxies the maximum velocity is equal to the plateau velocity
(this follows from the use of the approximation of a localized central
mass). If we consider the normal mass distribution (not localized at the
center), we see that the rotation velocity has the tendency to decrease
with diminishing radius, because of the diminishing effective mass near the
center.

\begin{figure}
\centerline{\psfig{figure=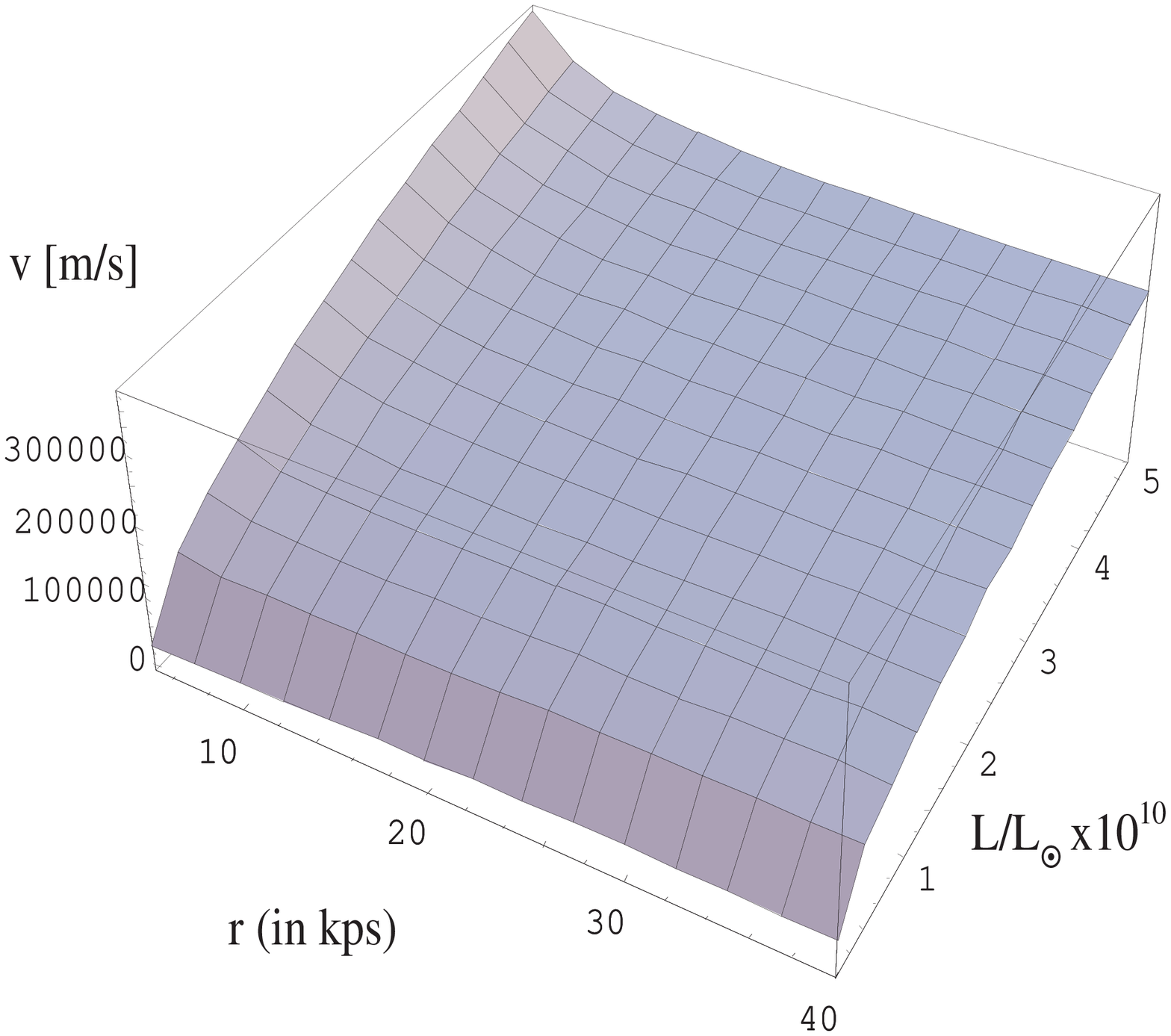,width=4in}}
\end{figure}

\centerline{\bf Fig.4}

\noindent From these two considerations, it is reasonable to expect to have
the maximum velocity difference equal to double the plateau velocity
for not very large galaxies. This indicates that the Tully-Fisher law is
true for smaller galaxies ($L < 5\times 10^{10}L_{\odot}$).

\section{The Dark Matter Problem Outside the Galaxy and Gravitational Lensing}

It is known that the nearest giant spiral galaxy M31 and our Galaxy form
the so-called Local Group. It is also known\cite{Tremaine} that the center
of M31 is approaching the center of the Galaxy at a velocity $\sim 119$
km/sec. This was an unexpected result, for most galaxies are moving
apart with the general Hubble law expansion. A natural explanation of this fact
was
given by Kahn and Woltjer\cite{Kahn}. They suggested that the relative
Hubble expansion of M31 and the Galaxy has been halted and reversed
by their mutual gravitational attraction. As was shown by them, the total
mass of the Local Group should be abnormally large, namely, the mass-to-light
ratio should be about $100\,(M/L)_{\odot}$. Such a big ratio is usually
explained by the dark matter hypothesis.

Let us consider whether it is possible to explain this phenomenon
without the dark matter hypothesis, using instead the gravity predictions of
NGT. Because the distance between M31 and the Galaxy is $\sim 700$ kpc,
we can safely use the point mass approximation. Nevertheless, we have to
consider our formulas as an approximation, because the masses of the galaxies
are not much different. Namely, the mass of M31 is about twice the mass
of the Galaxy. However, it is reasonable to apply our formalism, since
the disturbance due to the second mass is caused by the background metric
which is rather small. Moreover, if we apply our model for this
situation, the force of mutual attraction will be practically independent
of what we treat as the source and the test particle.

If we now calculate the gravitational attraction, we find that the
additional exponential force is vanishingly small. What is left contains the
renormalized gravitational constant, Eq.(\ref{renormgrav}), which contains the
factor:
\begin{equation}
1+\sqrt{\frac{M_0}{M}}=1+\sqrt{\frac{L_0}{L}}\sim 17.
\end{equation}
Thus, for the gravitational attraction we obtain:
\begin{equation}
a(r)=-\frac{G_0M^*}{r^2},
\end{equation}
where $M^*\sim 17M$ and from the ``observed" mass-to-light ratio:
\[
\frac{M^*}{L}\sim 100\biggl(\frac{M}{L}\biggr)_{\odot},
\]
we predict
\[
\frac{M}{L}\sim 6\biggl(\frac{M}{L}\biggr)_{\odot},
\]
which agrees with the estimated ratio for luminous matter without the
hypothesized dark matter.

Gravitational lensing experiments could give us new information about
the mass of galaxies\cite{Tyson}. We could estimate the value of the mass
expected
in such an experiment by using our model. Let us consider the lensing
effect produced by a galaxy. Because we consider a point mass approximation,
the distance $R$ between the galaxy center and the deflected light
ray must be greater than the size of the galaxy. For this case, we find
the angle of deflection $\Delta\phi$, obtained from a post-Newtonian expansion
of the field equations is\cite{Will}:
\begin{equation}
\Delta\phi=\frac{4G_0\biggl(1+\sqrt{M_0/M}\biggr)M}{c^2R},
\end{equation}
where $M$ is the mass of the galaxy.

We see that we anticipate a larger light deflection than is expected
to be produced by the luminous mass of the galaxy. This prediction is close
to the one that follows from the dark matter hypothesis.

\section{Terrestrial and Solar System Effects}

We have to consider the two questions:
{\smallskip\obeylines

1. What is the range of $r$ for which the approximate result,
Eq.(\ref{addacceleration}), is valid?

2. Can we estimate the corrections to the gravitational force in the solar
system and
for terrestrial experiments?
\smallskip}

The answer to the first question comes from Eq.(\ref{f2equation}). The
first correction comes from the exponential force, while the second
correction comes from the exact, static Wyman solution of the vacuum
NGT field equations\cite{Moffat1,Moffat2} valid in the long-range
approximation, $\mu\approx 0$. The region in which the
exponential force dominates is in the interval:
\[
\frac{1}{2}\biggl(\frac{r}{r_0}\biggr) > \frac{2G_{\infty}M}{c^2r}.
\]
For $r_0=25$ kpc, one has
\[
r > 0.2\biggl(\sqrt{\frac{M}{10^{10}M_{\odot}}}\biggr)\,kpc.
\]
Thus, we can use the exponential force formula as a correction up to
$r\geq 1$ kpc for a galaxy with $M=10^{10}M_{\odot}$, which
means that our approximation scheme is self-consistent.

Let us now address the second question. Choosing the constant $\lambda C$
obtained from our fits to the galaxies, we find that
\begin{equation}
\label{Cequation}
\lambda C\sim 10^{51}\sqrt{\frac{M}{M_{\odot}}}\,m^3.
\end{equation}
We see that the constant $C$ cannot be very large, since otherwise it will
lead to a big modification of the Schwarzschild solution of GR. If
we take the value of $\lambda C$ from Eq.(\ref{Cequation}), then we see that
the correction is much bigger than the Newtonian acceleration which is
unacceptable in our linear approximation. This means that we need to solve
the NGT field equations more
exactly in order to guarantee that the corrections are properly accounted for.
In particular, we must be able to renormalize the gravitational constant $G$
in the manner of Eq.(\ref{renormgrav}), within the small $r$ expansion of the
exact
NGT solution, to guarantee that the terrestrial and solar system corrections
are
small.

On the other hand, let us consider Eq.(\ref{totalaccel}) as an exact solution.
If we calculate the limit of this acceleration formula for small $r$, we get
the result:
\begin{equation}
\label{Newtcorrection}
\delta a(r)=\frac{a(r)-a_{\hbox{ngt}}(r)}{a_{\hbox{Newton}}(r)}
\approx \frac{1}{2}\sqrt{\frac{M_0}{M}}\biggl(\frac{r}{r_0}\biggr)^2.
\end{equation}
For $r_0\sim 25$ kpc, we find that for solar and terrestrial experiments,
$\delta a < 10^{-13}$. This deviation from the Newtonian
force law is too small to be detected with current experiments intended
to search for differences from Newtonian gravity, as well as a new
``fifth force" in nature (see, e.g.,\cite{Fischbach,Mostepanenko}).

\section{Conclusions}

{}From the linear weak field approximation to a generally covariant theory
of gravitation, which is free from ghost poles and tachyons, we have fitted
the flat rotation curves of galaxies and the Tully-Fisher law, $L\sim v^4$.
The fits were obtained for large as well as small galaxies. We also found
that it is reasonable to expect that we can explain some of the cluster
dynamics effects normally attributed to large amounts of dark matter,
although further work has to be carried out to improve on the point source
force law, before definite conclusions can be drawn about the fits to clusters
of galaxies.

We also showed that gravitational lensing predictions can be made on the
basis of the light deflection predicted at galaxy scales, since the
renormalized gravitational constant produces effects that simulate those
thought to be due to significant amounts of dark matter in galaxy halos.

The Yukawa-like force produced in the weak field limit of the NGT field
equations is an non-additive force that does not violate the weak
equivalence principle, and which appears to be consistent with solar
system and terrestrial experiments when considered as a self-consistent
solution of the NGT field equations.

\acknowledgements

This work was supported by the Natural Sciences and Engineering Research
Council of Canada. We thank L. Demopoulos and S. Tremaine for helpful and
stimulating discussions.

\end{document}